\documentclass[conference]{IEEEtran}

\IEEEoverridecommandlockouts

\usepackage{alltt}

\usepackage{tikz}
\usepackage{upquote}
\usetikzlibrary{decorations.pathreplacing}

\usepackage{graphicx}
\usepackage{comment}
\usepackage{subfig}
\usepackage{enumitem}
\usepackage[normalem]{ulem} 
\usepackage{amstext}
\usepackage{pifont}
\usepackage{MathDefs}
\usepackage{docmute}
\usepackage{circuitikz}
\usepackage{balance}
\usepackage{amsmath,amsfonts,amssymb,amsthm}
\usepackage{hyperref}
\hypersetup{
  colorlinks   = true, 
  urlcolor     = blue, 
  linkcolor    = blue, 
  citecolor   = red 
}
\usepackage{nomencl}
\makenomenclature
\usepackage{color, soul} 

\usepackage{tcolorbox} 
\newtcolorbox{Educ}[1]{
 title=#1,
  beamer, 
  colback=xlightblue,
  colframe=blue!30,
  fonttitle=\bfseries,
  left=1mm,
  right=1mm,
  top=1mm,
  bottom=1mm,
  middle=1mm,
  breakable,
}

\usepackage{url}
\usepackage{chapterbib}

\newtheorem{remark}{Remark}

\usepackage{algorithm}
\usepackage{algpseudocode} 

\usepackage{tabularx}
\newcolumntype{L}[1]{>{\raggedright\let\newline\\\arraybackslash\hspace{0pt}}m{#1}}
\newcolumntype{C}[1]{>{\centering\let\newline\\\arraybackslash\hspace{0pt}}m{#1}}
\newcolumntype{R}[1]{>{\raggedleft\let\newline\\\arraybackslash\hspace{0pt}}m{#1}}

\def\BibTeX{{\rm B\kern-.05em{\sc i\kern-.025em b}\kern-.08em
    T\kern-.1667em\lower.7ex\hbox{E}\kern-.125emX}}
\begin{document}

\title{Performance Analysis of Empirical Open-Circuit Voltage Modeling in Lithium Ion Batteries, \\Part-2: Data Collection Procedure

}

\author{\IEEEauthorblockN{P. Pillai, J. Nguyen, and B. Balasingam}
}

\maketitle

\begin{abstract}
This paper is the second part of a series of papers about empirical approaches to open circuit voltage (OCV) modeling and its performance comparison in lithium-ion batteries. 
The first part of the series \cite{slowOCVp1} introduced various sources of uncertainties in the OCV models and established a theoretical relationship between uncertainties and the performance of a battery management system.
In this paper, clearly defined approaches for low-rate OCV data collection are defined and described in detail. 
The data collection is designed with consideration to several parameters that affect the experimental time. Firstly, a more suitable method to fully charge the battery at different C-Rates is defined. 
Secondly, the OCV characterization following the full charge is described for various performance comparisons. 
Finally, optimal and efficient resistance estimation profiles are discussed.
From the voltage, current and time data recorded using the procedure described in this paper, the OCV-SOC relationship is characterized and its uncertainties are modeled in the third part \cite{slowOCVp3} of this series of papers. 
\end{abstract}

\begin{IEEEkeywords}
OCV-SOC modeling, 
OCV modeling, 
OCV-SOC characterization,
OCV characterization,
Li-ion batteries, 
state of charge
\end{IEEEkeywords}


\section{Introduction}

Battery management systems (BMS) rely on the open circuit voltage to state of charge (OCV-SOC) model to compute important diagnostic and prognostic states, such as the SOC, time to shout down (TTS) or remaining mileage, state of health (SOH), and the remaining useful life (RUL); these states are used in various battery management activities, such as, 
charging optimization,
cell balancing,
and
battery thermal management.

The OCV-SOC model  (also referred to as OCV characteristics, OCV-SOC characteristics, or the OCV curve) can be stored by the battery management system in one of the following three forms:
\begin{enumerate}[label=(\alph*)]
\item
{\bf An electrochemical model.}
The electrochemical model is a function that maps a particular OCV to its corresponding SOC value using several electrochemical properties of the battery 
\begin{align}
{\rm OCV} \triangleq V_\ro(s) = f_{\rm ec} (s,  c_0, \ldots, c_n) 
\end{align}
where the electrochemical OCV model parameters $ c_0, \ldots, c_n$ may include materials properties such as 
the standard redox potential of the given material, 
Boltzmann constant,
temperature, and the number of intercalated lithium ions in the anode \cite{birkl2015parametric}. 

\item 
{\bf A table.} 
The OCV-SOC table consists of data pairs spanning the entire SOC values ($s \in [0,1]$) and their corresponding OCV values. A sample OCV-SOC table is given below:

\begin{table}[h!]
\begin{center}
\begin{tabular}{|C{0.4 in}|C{0.8 in}|}
\hline 
SOC & OCV \\ \hline
$s_1=0$ & $v_1 = {\rm OCV_{min}}$ \\ \hline
$s_2$ & $v_2$ \\ \hline
\vdots & \vdots   \\ \hline
$s_n=1$ & $v_n = {\rm OCV_{max}}$ \\ \hline
\end{tabular}
\end{center}
\end{table}

where $ {\rm OCV_{min}}$ refers to the lowest OCV value (corresponding to the empty battery) and $ {\rm OCV_{max}}$ refers to the highest OCV value (corresponding to the fully charged battery). 
In empirical OCV modeling, the minimum and maximum OCV values,  $ {\rm OCV_{min}}$ and $ {\rm OCV_{max}}$ respectively, are custom defined to allow margins to safeguard the battery from over-charging and over-discharging.

\item
{\bf An empirical function. }
The empirical OCV-SOC model is a map of a particular OCV to its corresponding SOC value using several parameters as follows: 
\begin{align}
{\rm OCV} \triangleq V_\ro(s) = f_{\rm em} (s, k_0, \ldots, k_n) 
\end{align}
where the empirical OCV model parameters $k_0, \ldots, k_n$ are referred to in general as the OCV parameters or the ${\textit{k}}$-parameters in this paper. The difference between the empirical OCV model $f_{\rm em}$ and the electrochemical OCV model $f_{\rm ec}$ is that the empirical OCV model parameters, $k_0, k_1, \ldots, k_n$  are obtained based on data alone.
\end{enumerate}

Out of the above three approaches, the OCV-SOC table and the empirical OCV-SOC function are widely studied and adopted into BMS standards \cite{lai2018comparative}. 
These two OCV-SOC models only slightly differ based on how the parameters are stored:
an OCV-SOC table may need higher storage requirements (in the thousands of OCV-SOC pairs) compared to the OCV-SOC function which can store the OCV characteristics using less than ten ${\textit{k}}$-parameters \cite{pattipati2014open,pillai2022open}. 
However, the OCV-SOC table, when stored in very high resolution, is free of the modeling errors suffered by the empirical OCV function. 
It is also possible to strategically reduce the number of OCV-SOC pairs in a table \cite{sundaresan2022tabular}.

In this paper, empirical approaches to OCV modeling are studied.
Out of the two broad approaches (Galvanostatic Intermittent Titration Technique (GITT) and low-rate cycling approach), low-rate cycling is considered due to its potential to acquire high-resolution OCV-SOC data in a relatively shorter time \cite{christophersen2015battery}. 
In the low-rate cycling approach, the OCV cannot be directly measured due to the voltage drop;
during discharging, the terminal voltage is lower than the OCV;
during charging, the terminal voltage is higher than the OCV \cite{barai2019comparison}. 
For a low-rate constant current, the voltage drop can be written as 
\begin{align}
\text{voltage drop} = \text{current} \times \text{resistance} 
\end{align}
Since there is no control over the resistance, the current needs to be selected as small as possible such that the voltage drop can be reduced \cite{pattipati2014open}.  

\begin{figure}[h!]
\begin{center}
\subfloat[][True OCV-SOC curve. \label{fig:trueocvsoc}]
{\includegraphics[width=.9\columnwidth]{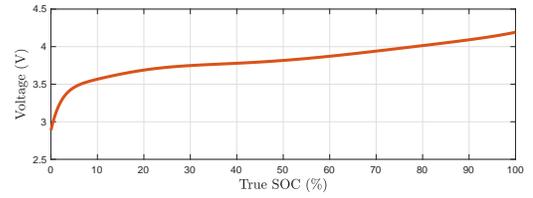}} \\
\subfloat[][BMS-defined OCV-SOC curve. \label{fig:bmsocvsoc}]
{\includegraphics[width=.9\columnwidth]{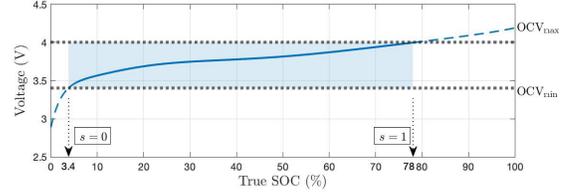}} \\
\subfloat[][SOC range with smaller drop. \label{fig:rel1ocvsoc}]
{\includegraphics[width=.9\columnwidth]{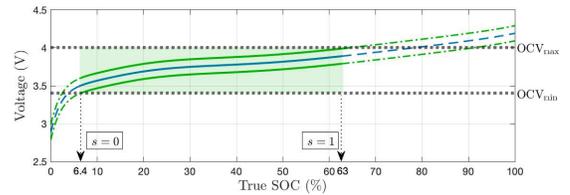}} \\
\subfloat[][SOC range with higher drop. \label{fig:rel2ocvsoc}]
{\includegraphics[width=.9\columnwidth]{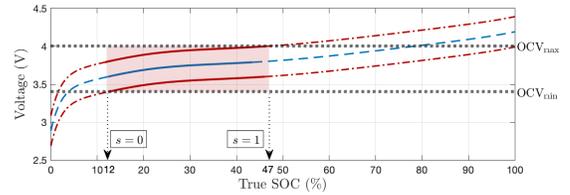}} \\
\subfloat[][Comparison of different SOC ranges.\label{fig:socranges}]
{\includegraphics[width=.9\columnwidth]{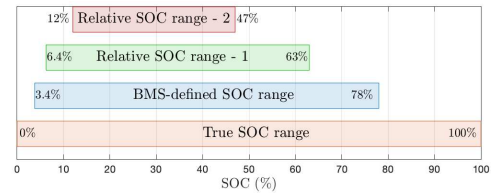}}
\caption{Relative SOC}
\label{fig:RelativeSOC}
\end{center}
\end{figure}

Figure \ref{fig:RelativeSOC} illustrates the dependency of the OCV-SOC curve on the C-Rate. 
In Figure \ref{fig:RelativeSOC}\subref{fig:trueocvsoc}, the solid orange line represents the true OCV-SOC curve of the battery; when the battery is completely empty (SOC = 0\%), the OCV is about 2.8V; and, when the battery is completely full  (SOC = 100\%), the OCV is about 4.2V. 
The goal of OCV characterization is to obtain data points representing the true OCV-SOC curve (orange line).
In the low-rate cycling approach to OCV modeling \cite{pillai2022open,pattipati2014open}, the battery is first discharged and then charged (or vice versa) using the same low C-Rate while continuously collecting the voltage and current data. 
When the charge and discharge portion of the voltage (corresponding to a certain SOC) is averaged, the resulting voltage is an accurate enough representation of the true OCV.

The voltage thresholds, ${\rm OCV_{min}}$ and ${\rm OCV_{max}}$, are set by the BMS to allow safety limits against over-discharging and overcharging, respectively; the low-rate cycling is performed within these limits (see Figure \ref{fig:RelativeSOC}\subref{fig:bmsocvsoc}).
The low-rate charging stops when the terminal voltage reaches ${\rm OCV_{max}}$; the relative SOC at this point is declared as ${\rm SOC}=1.$
The low-rate discharging stops when the terminal voltage reaches ${\rm OCV_{min}}$; relative the SOC at this point is declared as ${\rm SOC}=0.$
It can be noticed from Figure \ref{fig:RelativeSOC}\subref{fig:bmsocvsoc} that the BMS-defined SOC is within the range of true SOC.
During discharging, relative SOC reaches 0\% before the true SOC; during charging, BMS-defined SOC reaches 100\% before the true SOC. 

The goal in low-rate OCV modeling is to capture the OCV-SOC curve shown in Figure \ref{fig:RelativeSOC}\subref{fig:bmsocvsoc} between the OCV limits of ${\rm OCV_{min}}$ and ${\rm OCV_{max}}$. Alternatively, the captured OCV-SOC curve between the true SOC limits of 3.4\% and 78\% are now designated (relatively) as the BMS-defined SOC between 0\% and 100\%, respectively.

Figure \ref{fig:RelativeSOC}\subref{fig:rel1ocvsoc} shows how the low-rate OCV modeling approach attempts to approximate the true OCV-SOC model between the voltage thresholds defined in Figure \ref{fig:RelativeSOC}\subref{fig:bmsocvsoc}. The green lines in Figure \ref{fig:RelativeSOC}\subref{fig:rel1ocvsoc} show the terminal voltage that would be recorded during the low-rate OCV test---the terminal voltage is slightly above the true OCV during charging and below the true OCV during discharging. The magnitude of this voltage difference is approximately the same; hence, the true OCV can be obtained by averaging the voltages corresponding to charging and discharging.

It can be noticed in Figure \ref{fig:RelativeSOC}\subref{fig:rel1ocvsoc} that, due to the voltage drop, the entire OCV-SOC range cannot be captured in low-rate OCV modeling. According to the definition made in Figure \ref{fig:RelativeSOC}\subref{fig:bmsocvsoc}, the desired OCV-SOC range lies between true SOC of 3.4\% and 78\%. However, due to voltage drop, the resulting OCV-SOC model is restricted between true SOC of 6.4\% and 63\%. Figure \ref{fig:RelativeSOC}\subref{fig:rel2ocvsoc} shows that, as the voltage drop increases, the range of the relative OCV-SOC curve reduces. In this case, the red line indicates the terminal voltage that would be recorded during charging and discharging at a voltage drop higher than the one in Figure \ref{fig:RelativeSOC}\subref{fig:rel1ocvsoc}. Thus, the OCV-SOC curve here is only recorded for the true SOC limits of 12\% and 47\%. 

The BMS stores the OCV-SOC curve in one of the three forms discussed in (a)--(c) \cite{park2021complementary}. 
The SOC can be computed at any given time by measuring the OCV of the battery. 
However, to measure the OCV, the battery needs to be relaxed for several hours; this is virtually impossible in real-life applications such as EVs \cite{zhou2023precise}. 
To estimate the SOC in real-time numerous non-linear filtering techniques were proposed in the literature.
These approaches utilize the OCV-SOC function to model a state-space model as follows:
\begin{align}
s(t) &= f_{\rm cc} (s(t-1), Q, i(t), \Delta_t) \label{eq:ssm1} \\
v(t) &= f_{\rm em} (s(t), k_0, \ldots, k_n) + v_\rd(t) \label{eq:ssm2}
\end{align}
where
$s(t)$ denotes the SOC at time $t$,
$f_{\rm cc}(\cdot)$ denotes the Coulomb counting model \cite{Movassagh2021},
$Q$ denotes the battery capacity, 
$\Delta_t$ denotes the sampling time, 
and $v_\rd(t)$ denotes the voltage drop at time $t.$
The voltage drop of an active battery is modeled using the electrical equivalent circuit model (ECM) whose parameters need to be estimated \cite{bfg_part1,pillai2022optimizing}. 
Numerous non-linear filtering approaches were proposed for real-time SOC estimation using state-space models similar to the one in \eqref{eq:ssm1}-\eqref{eq:ssm2}; some examples of such filtering approaches include 
the extended Kalman filter (EKF) \cite{plett2004extended,yuan2022state,shi2022state}, 
the unscented Kalman filter (UKF) \cite{cui2021state,zhu2021novel}, and
particle filter (PF) \cite{khaki2021equivalent}. 

Little attention was paid in the literature about the possible uncertainties in the empirical OCV model $ f_{\rm em}(\cdot)$ in \eqref{eq:ssm2} and how such uncertainty can affect the performance of the SOC estimation and the BMS in general. 
For example, the battery capacity $Q$ in \eqref{eq:ssm1} fades over time and the OCV-SOC function $ f_{\rm em}(\cdot)$ is utilized by the BMS to estimate the battery capacity in real-time \cite{bfg_part1}. 
The goal of this paper, which is part of a series of papers, is to contribute to the understanding of empirical OCV-SOC modeling errors and ways to reduce them. 
In the first part \cite{slowOCVp1} of this series, uncertainty in $ f_{\rm em}(\cdot)$  due to several causes was identified and modeled. 
This paper, which is the second part of the series, presents data collection approaches to empirical OCV-SOC modeling. 
The third paper of the series \cite{slowOCVp3} utilizes the data collection approach presented in this paper to collect data from 16 commercially available identical battery cells.
Analysis of the data with regards to the performance metrics introduced in  \cite{slowOCVp1} is also presented in \cite{slowOCVp3}.

The remainder of this paper is organized as follows: 
Section \ref{sec:OCV-SOC-char}, details the theoretical background for data collection using the low-rate cycling method.
In Section \ref{sec:empOCVSOC} and \ref{sec:empOCVSOCtable}, details of developing an empirical OCV-SOC function and table using the collected data are summarized.
Section \ref{sec:R0-est} describes approaches to estimate the internal resistance of the battery. 
Section \ref{sec:DataCollectionProcedure} summarizes the data collection procedure and Section \ref{sec:conclusions} concludes the paper.

\section{Low-Rate OCV-SOC Data Collection}
\label{sec:OCV-SOC-char}

\begin{figure*}[h!]
\begin{center}
\subfloat[][DC equivalent circuit model of a battery.\label{fig:DC-ECM}]
{\begin{circuitikz}[american,scale=1, voltage dir = EF,line width=0.6pt]
\draw 
(0,0) to [battery1, v={EMF}] (0,3)
(0,3) -- (0.5,3)
(0.5,3) to[/tikz/circuitikz/bipoles/length=1 cm,cvsource, invert, v=$ h(k)$] (1.5,3)
(1.5,3) -- (2,3)
(2,3) to[/tikz/circuitikz/bipoles/length=0.8 cm,R=$ R_\Omega$] (3,3) 
(3,3) -- (3.75,3)
(4,3) to[short] (3.75,3)
(4,3) to[/tikz/circuitikz/bipoles/length=0.8 cm,R=$ R_{\rm SEI}$] (5.5,3)
(3.75,3) -- (3.75,2)
(3.75,2) -- (4,2)
(5.5,2) to[/tikz/circuitikz/bipoles/length=0.8 cm, cC, v^>=$ C_{\rm SEI}$] (4,2)
(5.5,2) -- (5.5,3)
(5.5,3) -- (5.75,3)
(6,3) to[short] (5.75,3)
(6,3) to[/tikz/circuitikz/bipoles/length=0.8 cm,R=$ R_{\rm CT}$] (7.5,3)
(5.75,3) -- (5.75,2)
(5.75,2) -- (6,2)
(7.5,2) to[/tikz/circuitikz/bipoles/length=0.8 cm,cC, v^>=$ C_{\rm DL}$] (6,2)
(7.5,2) -- (7.5,3)
(9,3) to[short, i_=${i(k)}$, o-] (7.5,3)
(9,3) to[open, v=${v(k)}$] (9,0) 
(9,0) to[short, o-] (0,0)
  ;
\end{circuitikz}
} \hspace{1cm}
\subfloat[][R-int equivalent circuit model.\label{fig:R-int}]
{\begin{circuitikz}[american,scale=1, voltage dir = EF,line width=0.6pt]
\draw
(0,0) to [battery1, v=EMF] (0,3)
(0,3) -- (0.5,3)
(0.5,3) to[/tikz/circuitikz/bipoles/length=1 cm,cvsource, invert, v=$ h(k)$] (1.5,3)
(1.5,3) -- (2,3)
(2,3) to[/tikz/circuitikz/bipoles/length=0.8 cm,R=$ R_0$] (3,3) 
(4,3) to[short, i_=${i(k)}$, o-] (3,3)
(4,3) to[open, v=${v(k)}$] (4,0)
(4,0) to[short, o-] (0,0)
  ;
\end{circuitikz}
}
\caption{Equivalent circuit models of a battery.}
\label{fig:ECM}
\end{center}
\end{figure*}
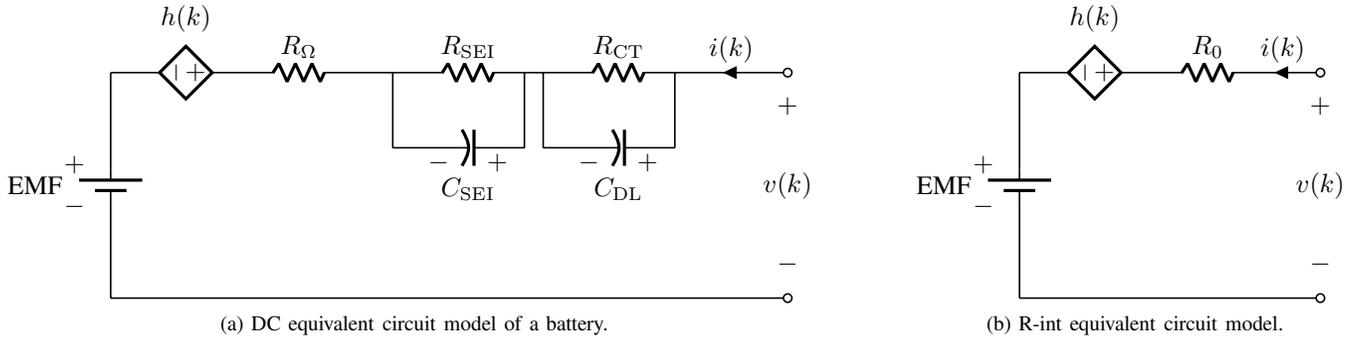

In the low-rate cycling approach to OCV modeling, one would continuously collect the OCV-SOC observation pairs $(s, V_o(s))$ that span the entire range of SOC (i.e., $s \in [0,1]$). 
However, the OCV is not directly measurable due to relaxation and hysteresis effects. A battery ECM consists of the OCV which represents its electromotive force (EMF), 
hysteresis effect $h(k)$,
internal Ohmic resistance $R_\Omega$, and the relaxation effect represented by two RC circuits in  Figure \ref{fig:ECM}.

A battery is considered active if it experiences a non-zero current $i(k)$ or if it has experienced a non-zero current $i(k)$ in the past few hours. 
Direct measurement of the OCV is not possible in an active battery due to the following effects:
\begin{enumerate}[label=(\roman*)]
\item
The Ohmic effect, due to the resistance $R_\Omega$ in the ECM.
The Ohmic effect can be easily mitigated by measuring $v(k)$ when $i(k)=0.$
\item
The relaxation effect, due to the capacitive (or RC) effect of the battery is shown in Figure \ref{fig:ECM}; 
this can be mitigated by resting the battery for several hours. 
This would mean taking one measurement for the OCV-SOC observation pair $(s, V_o(s))$ will take a couple of hours. 
Accurate OCV-SOC modeling will need hundreds of observation pairs spanning the entire region of SOC and OCV.  
\item
The hysteresis effect is quantifiable as a series voltage source (see Figure \ref{fig:ECM}). The hysteresis effect cannot be mitigated by resting the battery.  
It manifests as a bias in the direction of the voltage drop and its magnitude depends on the prior current and SOC of the battery. 
\end{enumerate}

\begin{remark}
Another way to mitigate the Ohmic, relaxation, and hysteresis effects is to estimate the voltages due to these three effects and then subtract them from $v(k).$
However, due to numerous uncertainties in modeling and parameter estimation, OCV characterization approaches, in general, seek to minimize these effects by controlling the current profile during the OCV experiment. 
\end{remark}

The low-rate cycling approach is designed to mitigate both the hysteresis and relaxation effects of the battery without substantially increasing the time taken for OCV characterization. 
\begin{enumerate}[label=(\alph*)]
\item
The relaxation effect is mitigated by employing a constant current to completely discharge and completely charge the battery; the capacitors of the battery ECM in Figure \ref{fig:ECM}\subref{fig:DC-ECM} will be saturated due to constant current and the resulting ECM will resemble the one shown in Figure  \ref{fig:ECM}\subref{fig:R-int}. 
\item
The hysteresis effect is mitigated by applying the same current to charge and discharge the battery; 
the hysteresis voltage will be in the opposite directions and will get eliminated when the entire charge-discharge data is used for parameter estimation. 
\item
In order to collect the data spanning the entire range of SOC (i.e. $s \in [0,1]$), corresponding to Figure \ref{fig:RelativeSOC}\subref{fig:bmsocvsoc}, the magnitude of the current will be kept as low as possible. 
In \cite{pattipati2014open}, the current was set at C/30 which required approximately 60 hours for data collection.  
In this paper, we will perform a systematic analysis to quantify the effect of the current rate on the resulting OCV curve.  
\end{enumerate}

Figure \ref{fig:ECM}\subref{fig:R-int} shows the R-int equivalent circuit model of a battery that is representative of battery ECM during low-rate OCV characterization. 
The R-int model is a reduced form of the complete ECM shown in Figure \ref{fig:ECM}\subref{fig:DC-ECM}; here, it is assumed that the current will always be constant and that the capacitors would be saturated as a result. 
The total resistance $R_0$ is related as follows to Figure \ref{fig:ECM}\subref{fig:DC-ECM}:
\begin{align}
R_0 = R_\Omega + R_{\rm SEI} + R_{\rm CT}
\label{eq:totalResistance}
\end{align}
where 
$R_\Omega$ denotes the Ohmic resistance,
$R_{\rm SEI}$ denotes the solid electrolyte resistance, and 
$R_{\rm CT}$ denotes the charge transfer resistance

%

\subsection{Data Collection for OCV Characterization}

The data collection for the OCV characterization needs to consider some constraints applicable to Li-ion batteries. 
First, the open circuit voltage of the battery is bounded by lower and upper limits as follows 
\begin{align}
 {\rm OCV_{min}}  <  {\rm EMF} \equiv V_o(s(k)) < {\rm OCV_{max}}
\end{align}
Correspondingly, the extreme values of the (relative) state of charge are defined as follows 
\begin{align}
s = 
\left\{ 
\begin{array}{cc}
0 & \text{for} \quad {\rm EMF} =  {\rm OCV_{min}}  \\
1 &  \text{for} \quad {\rm EMF} =  {\rm OCV_{max}} 
\end{array}
\right.
\end{align}
This SOC range corresponds to the scenario depicted in Figure \ref{fig:RelativeSOC}\subref{fig:bmsocvsoc}.
Second, for safety reasons, the terminal voltage needs to be kept within the allowable OCV limits, i.e., 
$ {\rm OCV_{min}}  <  v(k) < {\rm OCV_{max}}$.
This constraint limits the ability to collect data that reach the extreme points of SOC. 
Consequently, the current $i(k)$ needs to be as low as possible to achieve the SOC closer to its higher (i.e., $s(k)=1$) and lower  (i.e., $s(k)=0$) limits.  
In this paper, the OCV characterization will be performed at different C-Rates to quantify the effect of the C-Rate on the accuracy of the OCV curve.

\begin{figure}[h!]
\begin{center}
\subfloat[][Voltage data.]
{\includegraphics[width= 0.49\columnwidth]{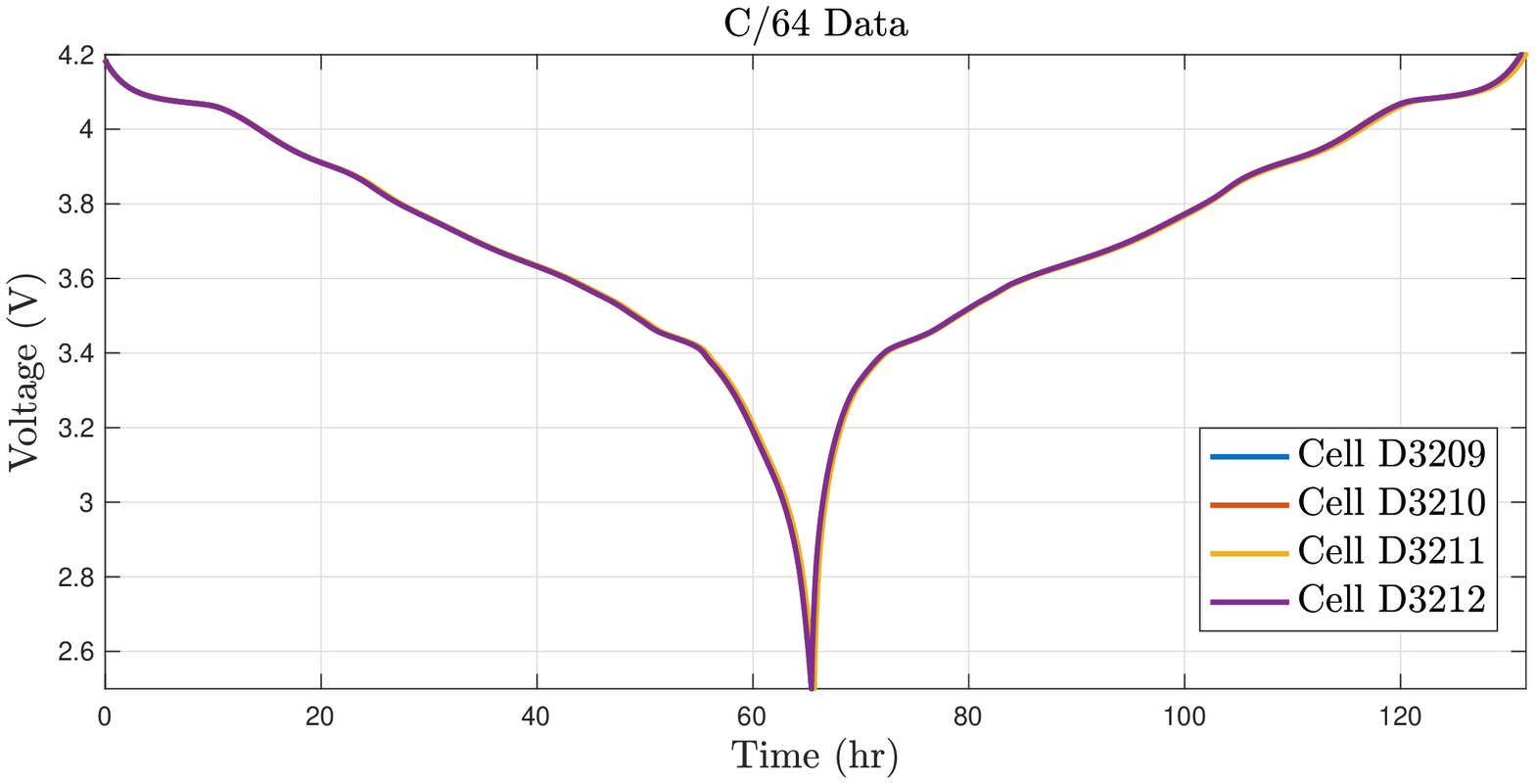}} 
\subfloat[][Current data.]
{\includegraphics[width= 0.49\columnwidth]{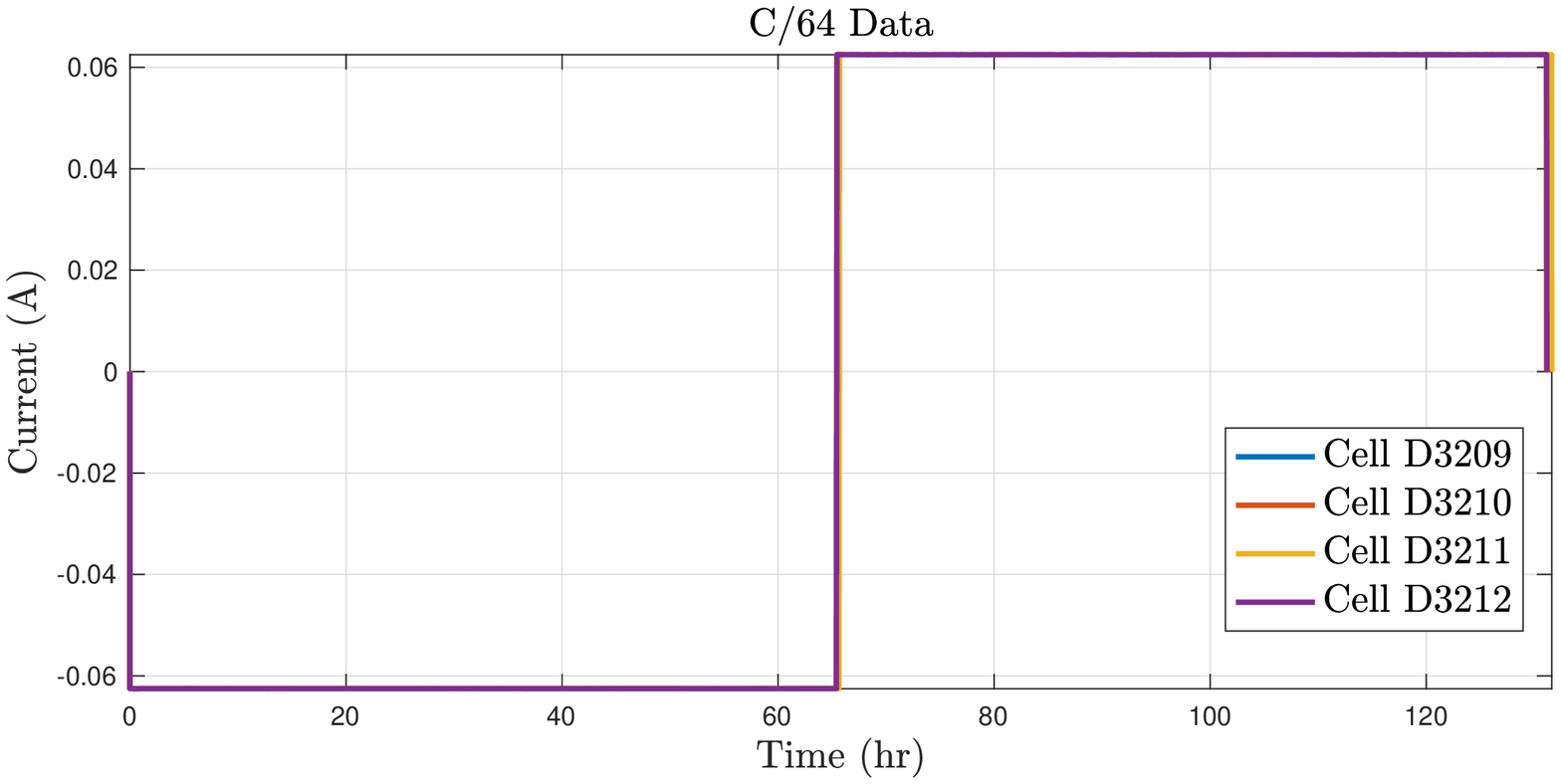}}
\caption{Current and voltage from a $C/64$ data collection.}
\label{fig:VIdataC64}
\end{center}
\end{figure}

Figure \ref{fig:VIdataC64} shows the current and voltage data collected for OCV characterization at the C/64 rate. 
The rated capacity of the battery is $4 \,\, {\rm Ah}$; this resulted in the current values of 
\begin{align} \nonumber
i(k) = I = 
\left\{
\begin{array}{ll}
I_c = 4/64 = 0.0625 \,\, {\rm A}& \text{charging} \\
I_d = -4/64 = -0.0625 \,\, {\rm A}& \text{discharging} 
\end{array}
\right.
\end{align}
Here, it must be noted that the current is selected to be 
\begin{enumerate}[label=(\alph*)]
\item
Constant so that the battery response reduces to the equivalent circuit model shown in Figure \ref{fig:ECM}\subref{fig:R-int}. 
\item
Very low so that the extreme values of SOC can be closely reached 
\item
The same during charging and discharging so that the non-linear effect of hysteresis can be cancelled during optimization
\end{enumerate}
It can also be noted from Figure \ref{fig:VIdataC64} that at $C/64$ rate, the total experimental time is close to 128 hours.

\subsection{Computing the Battery Capacity}

According to the OCV characterization data (shown for a C/64 experiment in Figure \ref{fig:VIdataC64} the battery was full ($s=1$) at the start of the experiment; and it was empty ($s=0$) at the end of discharge. 
This could only be an approximation due to the fact that there is a voltage drop at the amount of $v_d = I_c R_{0h} $ during charging and $v_d = I_d R_{0h} $ during discharging. 
The voltage drop prevents the battery OCV from reaching ${\rm OCV_{max}}$ during charging and ${\rm OCV_{min}}$ during discharging due to the strict cut-off voltage set for the terminal voltage at ${\rm OCV_{max}}$ and ${\rm OCV_{min}}$. To minimize this approximation error, the current $I$ needs to be as low as possible.

Thus, the capacity of the battery during discharging, denoted as $Q_{d}$  and charging, denoted as $Q_{c}$ can be defined as 
\begin{align}
Q_{c} =  I_{c} t_{c} \quad \text{and} \quad Q_{d} =  - I_{d} t_{d} \label{eq:Qd} 
\end{align}
where 
$I_{c}$ and $I_{d}$ are the charge and discharge currents respectively, and 
$t_{c}$ and $t_{d}$ are the charge and discharge times respectively.

\begin{remark}[Charging the battery prior to OCV characterization]
The battery must be charged to an appropriate level before the start of the experimental data collection shown in Figure \ref{fig:VIdataC64}. 
A detailed discussion on how to charge the battery before the start of the OCV characterization data collection is provided in Section \ref{sec:DataCollectionProcedure}. 
\end{remark}

\subsection{Computing the SOC}

The SOC values are needed for the OCV characterization data described in Table \ref{table:OCVchardata}. 
The Coulomb counting approach will be used to compute the SOC values for each sample as follows: 
\begin{align}
s(k+1) = s(k) + \frac{\Delta_k i(k)}{Q} \quad k = 1, 2, \ldots, m+n-1
\label{eq:CC}
\end{align}
where $\Delta_k = t(k+1) - t(k)$ is the time between $k^{\rm th}$ and $(k+1)^{\rm th}$ samples and 
\begin{align}
Q = 
\left\{
\begin{array}{cc}
Q_c & \text{if} \,\, i(k) > 0 \\
Q_d & \text{if} \,\,  i(k) < 0
\end{array}
\right.
\end{align}
Since the battery is considered full at the start, it will be assumed that $s(1)=1$.

Unlike the Coulomb counting approaches used in typical battery management systems \cite{Movassagh2021}, the uncertainties are minimal in the SOC values computed in \eqref{eq:CC}. 
The computed $s(k)$ values will be such that $s(m)=s(m+1)=0$ and $s(m+n)=1$ with the possibility of very small perturbations that may not affect the OCV characterization process nor will they affect the resulting OCV parameter values. 

\subsection{Required Data for OCV-SOC Characterization}

The voltage and current data for the OCV characterization will be collected according to the data collection procedure summarized in Section \ref{sec:DataCollectionProcedure}. 
Table \ref{table:OCVchardata} illustrates the data points that need to be collected for accurate estimation of the OCV parameters in \eqref{eq:vec-obs}. 
The first column indicates the mode of the battery: (-) indicates discharging and (+) indicates charging. 
During discharge mode, the battery is discharged from full to empty; during the subsequent charging mode, the battery is charged from empty to full. 
Accordingly, the SOC values during the discharge mode are $s(1)=1, \ldots, s(m) = 0$;  and the SOC values during the discharge mode are $s(m+1)=0, \ldots, s(m+n) = 1.$
Since the current is constant, the SOC difference between two adjacent data points will be a constant, i.e., $s(i+1) - s(i) = \Delta I / Q$ where $I$ is the current used in the OCV experiment and $Q$ is the battery capacity. The last column of Table \ref{table:OCVchardata} indicates scaled SOC values that are needed in some empirical models as elaborated in Section \ref{sec:scalingSOC}.
 
 \begin{table}[h!]
\caption{Required Data for OCV Characterization }
\label{table:OCVchardata}
\begin{center}
\begin{tabular}{|c|c|c|c|c|c|}
\hline 
Mode & Voltage & Current & SOC & SOC' \\ 
 &  $v(k)$& $i(k)$ &  $s(k)$  & $s'(k)$ \\ \hline
(-) & $v(1)$ & $i(1)$ & $s(1)$ & $s'(1)$   \\ \hline
\vdots  & \vdots & \vdots & \vdots & \vdots  \\ \hline
(-)  & $v(m)$ & $i(m)$ & $s(m)$ & $s'(m)$   \\ \hline
(+) & $v(m+1)$ & $i(m+1)$ & $s(m+1)$  & $s'(m+1)$    \\ \hline
\vdots  & \vdots & \vdots & \vdots  & \vdots \\ \hline
(+)  & $v(m+n)$ & $i(m+n)$ & $s(m+n)$  & $s'(m+n)$    \\ \hline
\end{tabular}
\end{center}
\end{table}%

\section{Empirical OCV-SOC Function}
\label{sec:empOCVSOC}

The goal in OCV-SOC modeling is to obtain the parameters ${\textit{k}}$, of the OCV curve. 
Several functions classified into linear, non-linear, hybrid, and tabular models were reported in the literature to model the OCV-SOC characteristics of a battery \cite{pattipati2014open,pillai2022open}. This section reviews a linear empirical model known as the Combined+3 model and the steps to compute the OCV parameters. It should be noted that the discussions presented in this section are relevant to other empirical OCV models also.

\subsection{OCV-SOC Model}
\label{sec:OCV-SOC-model}

The linear Combined+3 model for representing the OCV-SOC relationship is written as \cite{pattipati2014open}
\begin{align} 
\begin{aligned}
V_o(s) = 
& k_0 +k_1 s^{-1} +k_2s^{-2}+k_3s^{-3}+k_4s^{-4} + k_5s \\
&\quad \quad  +k_6 \ln(s)+k_7 \ln(1-s) 
\end{aligned}
\end{align}
In vector format, the Combined+3 model is written as
\begin{align} 
\begin{aligned}
V_o(s) = \bp_o(s(k))^T \bk_o
\label{eq:OCVmodel}
\end{aligned}
\end{align}
where
\begin{align}
\bp_o(s(k))^T &= \left[  1, \, 1/s, \, 1/s^2 , \, 1/s^3 , \, 1/s^4 , \,s , \, \ln(s), \, \ln(1-s)  \right] \\
\bk_o  &= [k_0, \,\, k_1, \,\, k_2,  \,\, k_3,  \,\, k_4,  \,\, k_5,  \,\, k_6,  \,\, k_7 ]^T
\label{eq:k-para}
\end{align}
The OCV parameters (or the ${\textit{k}}$-parameters alone represent the OCV-SOC curve for a given battery. 
Subsection \ref{sec:R0estimates} defines the approach to estimating the parameters defined in \eqref{eq:k-para}. 
The terminal voltage $ v(k)$ can be written as
\begin{equation}
v(k)= \underbrace{V_o(s(k))}_{\rm EMF} + \underbrace{ h(k) +i(k)R_0}_{v_d= i(k)R_{0h}}  \label{eq:obs-model}
\end{equation}
where
$V_o(s(k))$ is the open circuit voltage of the battery,
$h(k)$ is the voltage due to hysteresis effects,
$i(k)$ is the current through the battery, 
and
$R_0$ represents the internal resistance of the battery. 
The time index $(k)$ indicates that these quantities are collected over time---spanning complete discharge and charge of the battery. 

In \cite{pattipati2014open}, the hysteresis voltage is modelled approximately as a resistive element, i.e., 
\begin{align}
h(k) = i(k) R_h
\end{align}
The voltage observation \eqref{eq:obs-model} can then be written as
\begin{align}
    v(k)=V_o(s(k))+i(k)R_{0h}
\end{align}
where $R_{0h} = R_0 + R_h$.
In vector form, the observation model \eqref{eq:obs-model} can be written as 
\begin{align}
    v(k)= \underbrace
    {\begin{bmatrix} 
    \bp_o(s(k))^{T}&i(k) 
    \end{bmatrix}}_{\bp(s(k))^T} 
    \underbrace
    {\begin{bmatrix}
    \bk_o\\R_{0h}
    \end{bmatrix}}_{\bk} 
    \label{eq:vec-obs}
\end{align}

\subsection{Scaling}
\label{sec:scalingSOC}
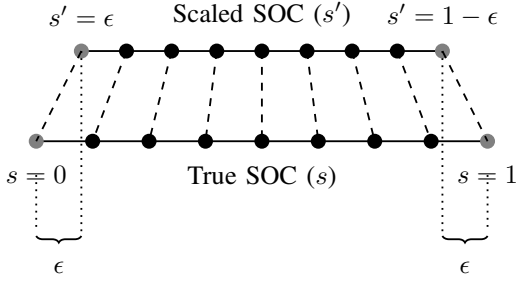
\begin{figure}[h!]
\begin{center}
\begin{circuitikz}[american,scale=.6, voltage dir = EF, line width=.7pt]
\draw 
(1,0)  node[circle, inner sep = -2pt,fill=gray]{} (1,0)
(1,0) -- (2,0)
(2,0)  node[circle, inner sep = -2pt,fill=black]{} (2,0)
(2,0) -- (3,0)
(3,0)  node[circle, inner sep = -2pt,fill=black]{} (3,0)
(3,0) -- (4,0)
(4,0)  node[circle, inner sep = -2pt,fill=black]{} (4,0)
(4,0) -- (5,0)
(5,0)  node[circle, inner sep = -2pt,fill=black]{} (5,0)
(5,0) -- (6,0)
(6,0)  node[circle, inner sep = -2pt,fill=black]{} (6,0)
(6,0) -- (7,0)
(7,0)  node[circle, inner sep = -2pt,fill=black]{} (7,0)
(7,0) -- (8,0)
(8,0)  node[circle, inner sep = -2pt,fill=black]{} (8,0)
(8,0) -- (9,0)
(9,0)  node[circle, inner sep = -2pt,fill=gray]{} (9,0)
(0,-2)  node[circle, inner sep = -2pt,fill=gray]{} (0,-2)
(0,-2) -- (1.25,-2)
(1.25,-2)  node[circle, inner sep = -2pt,fill=black]{} (1.25,-2)
(1.25,-2) -- (2.5,-2)
(2.5,-2)  node[circle, inner sep = -2pt,fill=black]{} (2.5,-2)
(2.5,-2) -- (3.75,-2)
(3.75,-2)  node[circle, inner sep = -2pt,fill=black]{} (3.75,-2)
(3.75,-2) -- (5,-2)
(5,-2)  node[circle, inner sep = -2pt,fill=black]{} (5,-2)
(5,-2) -- (6.25,-2)
(6.25,-2)  node[circle, inner sep = -2pt,fill=black]{} (6.25,-2)
(6.25,-2) -- (7.5,-2)
(7.5,-2)  node[circle, inner sep = -2pt,fill=black]{} (7.5,-2)
(7.5,-2) -- (8.75,-2)
(8.75,-2)  node[circle, inner sep = -2pt,fill=black]{} (8.75,-2)
(8.75,-2) -- (10,-2)
(10,-2)  node[circle, inner sep = -2pt,fill=gray]{} (10,-2)
(5,0) node[above, yshift=2mm] {Scaled SOC ($s'$)}
(5,-2) node[below, yshift=-2mm] {True SOC ($s$)}
(1,0) node[above, yshift=2mm] {$s' = \epsilon$}
(9,0) node[above, yshift=2mm] {$s' = 1 - \epsilon$}
(0,-2) node[below, yshift=-2mm] {$s = 0$}
(10,-2) node[below, yshift=-2mm] {$s = 1$};
\draw[dashed] (1,0) -- (0,-2) 
(9,0) -- (10,-2)
(2,0) -- (1.25,-2)
(3,0) -- (2.5,-2)
(4,0) -- (3.75,-2) 
(5,0) -- (5,-2)
(6,0) -- (6.25,-2) 
(7,0) -- (7.5,-2)
(8,0) -- (8.75,-2);
\draw[dotted]
(10,-2.75) -- (10,-4)
(9,0) -- (9,-4)
(0,-2.75) -- (0,-4)
(1,0) -- (1,-4);
\draw[decoration={brace},decorate]
(10,-4.1) -- node[below=6pt] {$\epsilon$} (9,-4.1);
\draw[decoration={brace},decorate]
(1,-4.1) -- node[below=6pt] {$\epsilon$} (0,-4.1);
\end{circuitikz}
\caption{
{\bf Linear scaling.} 
}
\label{fig:Scaling}
\end{center}
\end{figure}

Many OCV-SOC models including \eqref{eq:OCVmodel} consist of terms such as $1/s, \log(s)$, and $\log(1-s)$ that are not defined at either $s=0$ or at $s=1$. 
In order to avoid numerical instability, a linear scaling approach proposed in \cite{ahmed2020scaling} can be employed.  
Figure~\ref{fig:Scaling} illustrates a linear scaling with a scaling factor of $\epsilon$. 
The scaled SOC is written as
\begin{align}
 s' = (1-2\epsilon)s +\epsilon 
 \label{eq:s'}
 \end{align}
 where the value of $\epsilon$ needs to be selected based on the model; it was reported in \cite{ahmed2020scaling} that 
$\epsilon = 0.175$ yields minimal OCV modeling error in Combined model and its variants including the Combined+3 model.  

\subsection{OCV Parameter Estimation} 
\label{sec:R0estimates}

Based on the data points illustrated using Table \ref{table:OCVchardata}, the single observation model \eqref{eq:vec-obs} can be extended in matrix form for all the $(m+n)$ data points as follows
 \begin{align}
 \begin{bmatrix}
 v(1)\\ \vdots \\ v(m+n)
 \end{bmatrix}
 &=
  \begin{bmatrix}
\bp(s'(1))^T \\ \vdots \\ \bp(s'(m+n))^T 
 \end{bmatrix}
 \bk  \nonumber\\
 \bv & =  \bP \bk 
 \end{align}
 Given the observation vector $\bv$ and the matrix $\bP$, the parameter $\bk$ can be estimated using least squares approach, i.e.,
 \begin{equation}
   \hat \bk_{\rm LS} = (\bP^T \bP)^{-1} \bP^T \bv
 \label{eq:Kest}
\end{equation}
where 
\begin{align}
  \hat \bk_{\rm LS} = 
  \begin{bmatrix}
  \hat \bk_o \\ \hat R_{0h}
  \end{bmatrix} \label{eq:kLShat}
\end{align}

 Once the OCV parameters $\bk_o$ are estimated, the OCV for a given SOC, $s$, can be written as 
 \begin{align}
 \hat V_o(s) = \bp_o(s)^T \hat \bk_o 
 \label{eq:OCVestimate}
 \end{align}
 where $ \hat \bk_o$ denotes the estimate of $\bk_o.$

\section{Empirical OCV-SOC Table}
\label{sec:empOCVSOCtable}

\begin{figure}[h!]
\begin{center}
\subfloat[][Each `$\circ$' represents a SOC data point. All gray circles denote the endpoints and the points in black denote the intermediate SOC values.]
{\begin{circuitikz}[american,scale=0.6, voltage dir = EF, line width=.7pt]
\draw 
(1,0) node[circle, inner sep = -2pt, fill=gray]{}(1,0)
(1,0) -- (2,0)
(2,0) node[circle, inner sep = -2pt, fill=black]{}(2,0)
(2,0) -- (3,0)
(3,0) node[circle, inner sep = -2pt, fill=black]{}(3,0)
(3,0) -- (4,0)
(4,0) node[circle, inner sep = -2pt, fill=black]{}(4,0)
(4,0) -- (5,0)
(5,0) node[circle, inner sep = -2pt, fill=black]{} (5,0)
(5,0) -- (6,0)
(6,0) node[circle, inner sep = -2pt, fill=black]{} (6,0)
(6,0) -- (7,0)
(7,0) node[circle, inner sep = -2pt, fill=black]{}(7,0)
(7,0) -- (8,0)
(8,0) node[circle, inner sep = -2pt, fill=black]{}(8,0)
(8,0) -- (9,0)
(9,0) node[circle, inner sep = -2pt, fill=black]{} (9,0)
(9,0) -- (10,0)
(9,0) node[circle, inner sep = -2pt, fill=black]{} (9,0)
(10,0) -- (11,0)
(10,0) node[circle, inner sep = -2pt, fill=black]{} (11,0) 
(11,0) node[circle, inner sep = -2pt, fill=gray]{} (12,0)
(1.5,0) node[circle, inner sep = -2pt, fill=black]{}
(1.8,0) node[circle, inner sep = -2pt, fill=black]{}
(2,0) node[circle, inner sep = -2pt, fill=black]{}
(3.3,0) node[circle, inner sep = -2pt, fill=black]{}
(3.8,0) node[circle, inner sep = -2pt, fill=black]{}
(4.9,0) node[circle, inner sep = -2pt, fill=black]{}
(5.6,0) node[circle, inner sep = -2pt, fill=black]{}
(6.1,0) node[circle, inner sep = -2pt, fill=black]{}
(6.3,0) node[circle, inner sep = -2pt, fill=black]{}
(7.5,0) node[circle, inner sep = -2pt, fill=black]{}
(7.9,0) node[circle, inner sep = -2pt, fill=black]{}
(8.1,0) node[circle, inner sep = -2pt, fill=black]{}
(8.7,0) node[circle, inner sep = -2pt, fill=black]{} 
(8.9,0) node[circle, inner sep = -2pt, fill=black]{}
(9.3,0) node[circle, inner sep = -2pt, fill=black]{}
(9.7,0) node[circle, inner sep = -2pt, fill=black]{}
(10.2,0) node[circle, inner sep = -2pt, fill=black]{}
(10.3,0) node[circle, inner sep = -2pt, fill=black]{}
(10.8,0) node[circle, inner sep = -2pt, fill=black]{};
\node at (6,-0.7) (A) {OCV-SOC data};
\draw 
(1,-2) node[circle, inner sep = -2pt, fill=gray]{}(1,-2)
(1,-2) -- (2,-2)
(2,-2) node[circle, inner sep = -2pt, fill=black]{}(2,-2)
(2,-2) -- (3,-2)
(3,-2) node[circle, inner sep = -2pt, fill=black]{}(3,-2)
(3,-2) -- (4,-2)
(4,-2) node[circle, inner sep = -2pt, fill=black]{}(4,-2)
(4,-2) -- (5,-2)
(5,-2) node[circle, inner sep = -2pt, fill=black]{} (5,-2)
(5,-2)-- (6,-2)
(6,-2)node[circle, inner sep = -2pt, fill=black]{} (6,-2)
(6,-2)-- (7,-2)
(7,-2)node[circle, inner sep = -2pt, fill=black]{}(7,-2)
(7,-2)-- (8,-2)
(8,-2)node[circle, inner sep = -2pt, fill=black]{}(8,-2)
(8,-2)-- (9,-2)
(9,-2) node[circle, inner sep = -2pt, fill=black]{} (9,-2)
(9,-2)-- (10,-2)
(10,-2) node[circle, inner sep = -2pt, fill=black]{} (11,-2)
(10,-2)-- (11,-2)
(11,-2) node[circle, inner sep = -2pt, fill=gray]{} (12,-2);
\node at (6,-2.7) (A) {OCV-SOC table}; 
\vspace{1cm}
\draw
(1,0) node[above, yshift=2mm] {$s = 0$}
(10.8,0) node[above, yshift=2mm] {$s = 1$}
(1,-0.3) node[below, yshift=2mm] {$1^{st}$ point}
(10.8,-0.3) node[below, yshift=2mm] {$(m+n)^{th}$ point};
\draw
(1,-2) node[above, yshift=2mm] {$s = 0$}
(10.8,-2) node[above, yshift=2mm] {$1^{st}$ point}
(1,-2.3) node[below, yshift=2mm] {$1$}
(10.8,-2.3) node[below, yshift=2mm] {$N^{th}$ point};
\end{circuitikz}} \\
\subfloat[][Empirical OCV-SOC table for a sample OCV-SOC data.]
{\includegraphics[width= 0.9\columnwidth]{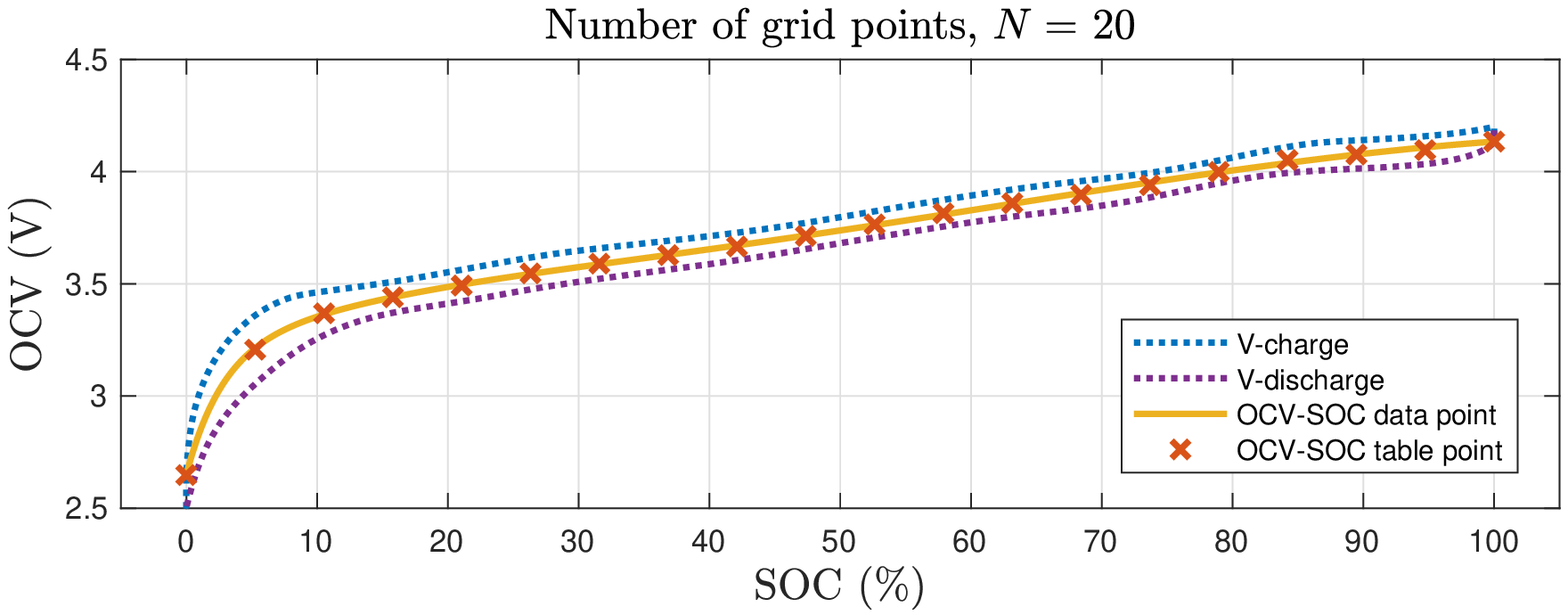}}
\caption{\bf Empirical OCV-SOC table.}
\label{fig:empOCVSOCtableplot}
\end{center}
\end{figure}

Consider the OCV-SOC data summarized in Table \ref{table:OCVchardata}.
The number of data points depends on the sampling time and the charging-discharging times, $t_c$ and $t_d$, respectively. The number of samples during the charging and discharging modes, $n$ and $m$ respectively, may also differ. In typical low-rate OCV experiments \cite{pattipati2014open}, where the sampling time is 1 minute, $m$ and $n$ will be in thousands. From the raw data shown in Table \ref{table:OCVchardata}, OCV-SOC data can be obtained as illustrated in Figure \ref{fig:empOCVSOCtableplot}.
Here, the SOC is split into $N$ points between 0 and 1. The corresponding OCV values are then located using linear interpolation. 

\section{Internal Resistance Estimation}
\label{sec:R0-est}

The OCV parameter estimation approach in \eqref{eq:Kest} is designed to estimate the equivalent resistance $R_{0h} =R_0 + R_h$ that is modeled to represent combined voltage drop due to hysteresis and electric resistance. 
In this section, an approach based on \cite{pillai2022optimizing}  is briefed to separately estimate the internal resistance $R_0$ of the battery. 

\subsection{Theoretical Background }
It was shown in \cite{pillai2022optimizing} that the battery internal resistance can be optimally estimated by applying a pulse charge/discharge current of equal amplitudes. 
Let us consider the R-int equivalent circuit model shown in Figure \ref{fig:ECM}\subref{fig:R-int}. 
Here, the voltage $E(k)$ represents the sum of EMF and the hysteresis voltage shown in Figure \ref{fig:ECM}\subref{fig:R-int}. 
The measured voltage across the battery terminals at time $(k)$ is written as 
\begin{align}
z_v(k) &= v(k) + n(k) =  i(k) R_0 +  E(k) + n(k) 
\end{align}
where $n(k)$ is the voltage measurement noise that is assumed to be zero-mean with standard deviation $\sigma.$

Now, considering $L$ consecutive observations of measured terminal voltage,
\begin{align}
\begin{aligned}
z_v(1) &=  i(1) R_0 +  E(1) + n(1) \\
z_v(2) &=  i(2) R_0 +  E(2) + n(2) \\
& \vdots \quad \quad \quad \vdots  \quad \quad \quad \vdots  \quad \quad \quad \vdots  \\
z_v(L) &=  i(L) R_0 +  E(L) + n(L) \\
\end{aligned}
\label{eq:obs}
\end{align}

The observations in \eqref{eq:obs} can be rewritten in vector format as 
\begin{align}
\bz = \bH \bx + \bn \label{eq:vecobs}
\end{align}
where 
\begin{align}
\bz=
\begin{bmatrix}
z_v(1)\\
z_v(2)\\
\vdots \\
z_v(L)
\end{bmatrix}, \,\,
\bH=
\begin{bmatrix}
i(1) & 1\\
i(2) & 1\\
\vdots & \vdots \\
i(L) & 1
\end{bmatrix}, \,\,
\bx = 
\begin{bmatrix}
R_0\\
E
\end{bmatrix}, \,\,
\bn=
\begin{bmatrix}
n(1)\\
n(2)\\
\vdots \\
n(L)
\end{bmatrix}
\end{align}
and it is assumed that 
\begin{align}
E(1) = E(2) = \ldots = E(L) = E
\label{eq:assumption}
\end{align}
The current profile needs to be selected in a way that the above assumption can be (completely or approximately) satisfied. This will be discussed later in this section.

The least-square estimate of $\bx$ in \eqref{eq:vecobs} is a vector of estimated resistance $\hat R_0$ and open-circuit voltage $\hat E$, and is given as
\begin{align}
\hat \bx_{\rm LS} = \begin{bmatrix}
\hat R_0 \\
\hat E
\end{bmatrix}
= \left( \bH^T \bH \right)^{-1} \bH^T \bz 
\label{eq:xLS}
\end{align}

From \cite{pillai2022optimizing}, it can also be shown that the Cramer-Rao lower bound (${\rm CRLB}$) of the estimate under the above assumptions is
\begin{align}
{\rm CRLB} = \left( \bH^T \bSigma^{-1} \bH \right)^{-1} = \sigma^2  \left( \bH^T \bH \right)^{-1}
\end{align}
which is a $2 \times 2$ matrix.
The ${\rm CRLB}$ is the theoretical lower bound on the covariance of the estimation error of $\hat \bx_{\rm LS}$.
The $(1,1)$ element of the above matrix represent the estimation error variance of the resistance, from \cite{pillai2022optimizing} is given as
\begin{align}
\sigma_{R_0}^2 
= E \left( (\hat R_0 -R_0 )^2 \right) = \frac{\sigma^2}{  \sum_{k=1}^{L} i(k)^2 - \frac{1}{L} \left(\sum_{k=1}^{L} i(k) \right)^2} 
\label{eq:E(R0err)}
\end{align}

\def\efficientEST{For an efficient estimator the variance of its estimation error is the same as the ${\rm CRLB}$.}

It can be shown that under the stated assumptions, the resistant estimator in \eqref{eq:xLS} efficient, i.e., the variance of the estimation error would be the same as the ${\rm CRLB}$ \cite{bar2004estimation}. 

\begin{remark}[Efficient vs. optimal estimator]
The resistance estimator defined in \eqref{eq:xLS} is proved to be efficient (as all least square estimators are under typical assumptions). 
An optimal estimator (introduced in \cite{pillai2022optimizing}) selects the current profiles $i(k)$ in such a way that the ${\rm CRLB}$ \eqref{eq:E(R0err)} can be minimized to its lowest possible value. 
\end{remark}

\subsection{Resistance Estimation Profiles}
\label{sec:R-est-discharge-pulse}

In this section, two types of current profiles are discussed for internal resistance estimation. 
The first one shown in Figure \ref{fig:R0estPulse1}, details a current profile that is suitable when the battery is closer to full and the second profile in Figure \ref{fig:R0estPulse2} is for when the battery is close to empty. 
Following these details, a current profile that is suitable when the battery SOC is in the mid-range is given in Figure \ref{fig:R0estPulse3}; an advantage of this profile is that it is designed to minimize the ${\rm CRLB}$ derived in \eqref{eq:E(R0err)} (i.e., this profile is designed for optimal estimation of the resistance).

\subsubsection{Fully charged battery}
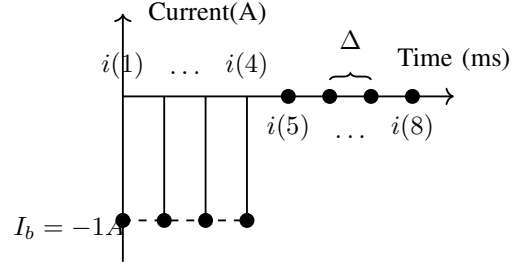
\begin{figure}[h!]
\begin{center}
\begin{circuitikz}[american,scale=.55, voltage dir = EF, line width=.7pt]
\draw[->]
(0,-4) -- (0,2);
\draw[->]
(0,0) -- (8,0);
\draw
(1,0) -- (1,-3)
(2,0) -- (2,-3)
(3,0) -- (3,-3)
(0,-3)  node[circle, inner sep = -2pt,fill=black]{} (0,-3)
(1,-3)  node[circle, inner sep = -2pt,fill=black]{} (1,-3)
(2,-3)  node[circle, inner sep = -2pt,fill=black]{} (2,-3)
(3,-3)  node[circle, inner sep = -2pt,fill=black]{} (3,-3)

(4,0)  node[circle, inner sep = -2pt,fill=black]{} (4,0)
(5,0)  node[circle, inner sep = -2pt,fill=black]{} (5,0)
(6,0)  node[circle, inner sep = -2pt,fill=black]{} (6,0)
(7,0)  node[circle, inner sep = -2pt,fill=black]{} (7,0);
\draw[dashed]
(0,-3) -- (3,-3);
\draw[decoration={brace},decorate]
(5,0.4) -- node[above=8pt] {$\Delta$} (6,0.4);
\draw
(0,-0.2) node[above=6pt] {$i(1)$}
(1.5,-0.2) node[above=8pt] {$\ldots $}
(3,-0.2) node[above=6pt] {$i(4)$}

(4,0.2) node[below=6pt] {$i(5)$}
(5.5,0.2) node[below=14pt] {$\ldots $}
(7,0.2) node[below=6pt] {$i(8)$}

(-1.3,-2.3) node[below=6pt] {$I_b = -1A$}
(8,0) node[above=6pt] {Time (ms)}
(0,2) node[right=6pt] {Current(A)};
\end{circuitikz}
\caption{Discharge pulse for resistance estimation. This pulse is suitable to estimate the resistance of a fully charged battery. \label{fig:R0estPulse1}}
\end{center}
\end{figure}

Figure \ref{fig:R0estPulse1} shows a current pulse that can be used to estimate the internal resistance of a (close to) fully charged battery.  
Here, only the discharging current is used because the charging current cannot be applied to an already fully charged battery to prevent the terminal voltage from exceeding the ${\rm OCV_{max}}$ threshold. 

For the current profile given in Figure \ref{fig:R0estPulse1}, the current values are 
\begin{align}
i(k) = 
\left\{ 
\begin{array}{cl}
-I_b& \text{for $k=1,\ldots, 4$}\\
0 &  \text{for $k=5,\ldots, 8$}
\end{array}
\right.
\label{eq:i(k)values}
\end{align}
According to the analysis presented in \cite{pillai2022optimizing}, the sampling time $\Delta$ needs to be as low as possible.
Based on the values for $i(k)$ in \eqref{eq:i(k)values}, the resistance estimation error variance is 
\begin{align}
\sigma_{R_0}^2 = \frac{\sigma^2}{  \sum_{k=1}^{8} i(k)^2 - \frac{1}{8} \left(\sum_{k=1}^{8} i(k) \right)^2} = \frac{\sigma^2}{2}
\label{eq:E(R0err1)}
\end{align}

\begin{remark}
The assumption \eqref{eq:assumption} is violated by measurements $6,7,8.$
The voltage difference between $E(1)$ and $E(8)$ can be computed to verify that the difference is negligibly small, approximately, $4.1917-4.187 =0.0047$V (see below). 
\begin{table}[h!]
\begin{center}
\begin{tabular}{|c|c|}
\hline 
SOC & OCV \\ \hline
$s(1)=1$ & $E(1)=4.1917$V \\ \hline
$s(5) =  0.9962$ & $E(5)=4.1870$V \\ \hline
$s(8) = 0.9962$ & $E(8)=4.1870$V \\ \hline
\end{tabular}
\end{center}
\end{table}
\end{remark}

\subsubsection{Empty battery}
\begin{figure}[h!]
\begin{center}
\begin{circuitikz}[american,scale=.55, voltage dir = EF, line width=.7pt]
\draw[->]
(0,-2) -- (0,4);
\draw[->]
(0,0) -- (8,0);
\draw
(1,0) -- (1,3)
(2,0) -- (2,3)
(3,0) -- (3,3)
(0,3)  node[circle, inner sep = -2pt,fill=black]{} (0,3)
(1,3)  node[circle, inner sep = -2pt,fill=black]{} (1,3)
(2,3)  node[circle, inner sep = -2pt,fill=black]{} (2,3)
(3,3)  node[circle, inner sep = -2pt,fill=black]{} (3,3)

(4,0)  node[circle, inner sep = -2pt,fill=black]{} (4,0)
(5,0)  node[circle, inner sep = -2pt,fill=black]{} (5,0)
(6,0)  node[circle, inner sep = -2pt,fill=black]{} (6,0)
(7,0)  node[circle, inner sep = -2pt,fill=black]{} (7,0);
\draw[dashed]
(0,3) -- (3,3);
\draw[decoration={brace},decorate]
(5,0.4) -- node[above=8pt] {$\Delta$} (6,0.4);
\draw
(0,0.2) node[below=6pt] {$i(1)$}
(1.5,0.2) node[below=14pt] {$\ldots $}
(3,0.2) node[below=6pt] {$i(4)$}

(4,0.2) node[below=6pt] {$i(5)$}
(5.5,0.2) node[below=14pt] {$\ldots $}
(7,0.2) node[below=6pt] {$i(8)$}

(-1.2,2.3) node[above=6pt] {$I_b = 1A$}
(8,0) node[above=6pt] {Time (ms)}
(0,4) node[right=6pt] {Current(A)};
\end{circuitikz}
\caption{Charge pulse for resistance estimation. This pulse is suitable to estimate the resistance of an empty battery. \label{fig:R0estPulse2}}
\end{center}
\end{figure}
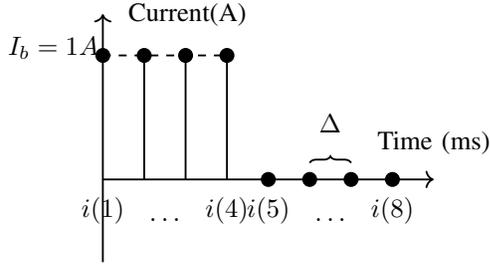

Figure \ref{fig:R0estPulse2} shows a current pulse that can be used to estimate the internal resistance of a (close to) empty battery.  
Here, only the charging current is used because the discharging current cannot be applied to an already empty battery to prevent the terminal voltage from exceeding the ${\rm OCV_{min}}$ threshold. 

For the current profile given in Figure \ref{fig:R0estPulse2}, the current values are 
\begin{align}
i(k) = 
\left\{ 
\begin{array}{cl}
I_b & \text{for $k=1,\ldots, 4$}\\
0 &  \text{for $k=5,\ldots, 8$}
\end{array}
\right.
\label{eq:i(k)values}
\end{align}
The resistance estimation error variance can be calculated to be the same as the variance given in \eqref{eq:E(R0err1)}.

\begin{remark}
The assumption \eqref{eq:assumption} is violated by measurements $1,2, 3, 4.$
The voltage difference between $E(1)$ and $E(8)$ can be computed to verify that the difference is quite small, approximately, $2.886-2.9839 =-0.0979$V (see below).
\begin{table}[h!]	
\begin{center}
\begin{tabular}{|c|c|}
\hline 
SOC & OCV \\ \hline
$s(1)=0$ & $E(1)=2.8860$V \\ \hline
$s(5) =  0.0039$ & $E(5)=2.9839$V \\ \hline
$s(8) = 0.0039$ & $E(8)=2.9839$V \\ \hline
\end{tabular}
\end{center}
\end{table}
\end{remark}

\subsection{Optimized Resistance Estimation Profile}
\label{sec:R-est-oprimized-pulse}

When the battery SOC is in the mid-range, it is safe to apply both charging and discharging current pulses for resistance estimation. 
In this section, based on \cite{pillai2022optimizing}, a current profile that can minimize the CRLB in \eqref{eq:E(R0err)} is presented.

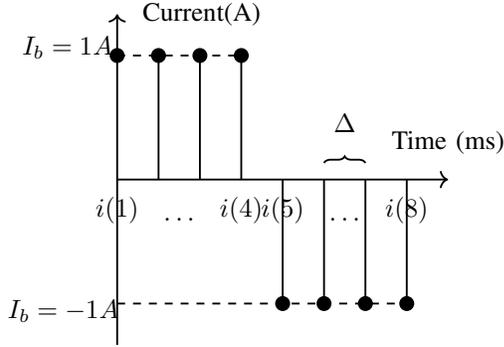
\begin{figure}[h!]
\begin{center}
\begin{circuitikz}[american,scale=.55, voltage dir = EF, line width=.7pt]
\draw[->]
(0,-4) -- (0,4);
\draw[->]
(0,0) -- (8,0);
\draw
(1,0) -- (1,3)
(2,0) -- (2,3)
(3,0) -- (3,3)
(4,0) -- (4,-3)
(5,0) -- (5,-3)
(6,0) -- (6,-3)
(7,0) -- (7,-3)
(0,3)  node[circle, inner sep = -2pt,fill=black]{} (0,3)
(1,3)  node[circle, inner sep = -2pt,fill=black]{} (1,3)
(2,3)  node[circle, inner sep = -2pt,fill=black]{} (2,3)
(3,3)  node[circle, inner sep = -2pt,fill=black]{} (3,3)

(4,-3)  node[circle, inner sep = -2pt,fill=black]{} (4,0)
(5,-3)  node[circle, inner sep = -2pt,fill=black]{} (5,0)
(6,-3)  node[circle, inner sep = -2pt,fill=black]{} (6,0)
(7,-3)  node[circle, inner sep = -2pt,fill=black]{} (7,0);
\draw[dashed]
(0,3) -- (3,3)
(0,-3) -- (7,-3);
\draw[decoration={brace},decorate]
(5,0.4) -- node[above=8pt] {$\Delta$} (6,0.4);
\draw
(0,0.2) node[below=6pt] {$i(1)$}
(1.5,0.2) node[below=14pt] {$\ldots $}
(3,0.2) node[below=6pt] {$i(4)$}

(4,0.2) node[below=6pt] {$i(5)$}
(5.5,0.2) node[below=14pt] {$\ldots $}
(7,0.2) node[below=6pt] {$i(8)$}

(-1.2,2.3) node[above=6pt] {$I_b = 1A$}
(-1.3,-2.3) node[below=6pt] {$I_b = -1A$}
(8,0) node[above=6pt] {Time (ms)}
(0,4) node[right=6pt] {Current(A)};
\end{circuitikz}
\caption{Pulse for optimized resistance estimation. \label{fig:R0estPulse3}}
\end{center}
\end{figure}

Figure \ref{fig:R0estPulse3} shows one cycle of the optimal current profile \cite{pillai2022optimizing}. 
It can be easily verified that the more cycles, the lower the estimation error. 
Since the battery SOC will not change due to this current profile, more cycles can be applied to lower the estimation error in practical systems where the sensor measurement noise is significant. 

For one cycle of the current profile given in Figure \ref{fig:R0estPulse3}, the current values are 
\begin{align}
i(k) = 
\left\{ 
\begin{array}{cl}
I_b & \text{for $k=1,\ldots, 4$}\\
-I_b &  \text{for $k=5,\ldots, 8$}
\end{array}
\right.
\label{eq:i(k)values2}
\end{align}
The above profile can be repeated for multiple cycles in order to reduce estimation error.

Considering $m$ consecutive cycles of the current profile shown in Figure \ref{fig:R0estPulse3}, the resistance estimation error variance is calculated as
\begin{align}
\sigma_{R_0}^2 = \frac{\sigma^2}{  \sum_{k=1}^{8m} i(k)^2 } =\frac{\sigma^2}{8 m I_b^2}
\end{align}
For similar values used in Section \ref{sec:R-est-discharge-pulse} (i.e., for $I_b=1$ and $m=1$ cycle), the estimation error variance is $\sigma_{R_0}^2 = {\sigma^2/}{8}$.

\begin{remark}
The error difference between the approaches presented in Section \ref{sec:R-est-discharge-pulse} and Section \ref{sec:R-est-oprimized-pulse} are very low when scientific grade measurement systems are used in laboratory settings.
For example, the equipment used in this series of papers has a very low measurement error standard deviation of approximately $\sigma = 0.2 \,\, {\rm mV}.$ Details of the experimental setup are available in \cite{slowOCVp3}.
Correspondingly, the resistance estimation error stand deviations are $\sqrt{\sigma^2/2}= 0.14 \,\, {\rm m} \Omega$ and $\sqrt{\sigma^2/8}= 0.07 \,\, {\rm m} \Omega$, respectively, for current profiles discussed in Section \ref{sec:R-est-discharge-pulse} and Section \ref{sec:R-est-oprimized-pulse}, respectively. 
\end{remark}

 Once the voltage and current data are collected, the internal resistance can be estimated using the approach summarized in Section \ref{sec:R0-est}. 
As mentioned before, internal resistance estimation is not required for OCV parameter estimation. 
In this paper, internal resistance estimation is performed to discuss the validity of the model presented in Section \ref{sec:OCV-SOC-model}. 

\subsection{Computing the Hysteresis Voltage}

The voltage drop in a battery equivalent circuit model consists of the voltage across the RC equivalent circuit (which is used to model the relaxation effect of the battery) and the hysteresis. 
Thus, from \eqref{eq:kLShat}, the total resistance can be recovered from the OCV parameters as follows: 
\begin{align}
\hat R_{0h} = \hat \bk_{\rm LS}(9)
\end{align}
where $\hat \bk_{\rm LS}$ denotes the parameters of the Combined+3 model that is used to model the empirical OCV. 
From this, an estimate of the hysteresis equivalent resistance can be obtained as  
\begin{align}
 \hat R_h = \hat R_{0h} - \hat R_0 
\end{align}
where $\hat R_{0h} $ denotes the estimate of $R_0$ obtained using the least squares approach described in Section \ref{sec:R0estimates};
and $ \hat R_0 $ denotes the estimate of $R_0$ for each battery described in Section \ref{sec:R0-est}.

Now, the estimates of hysteresis can be obtained using the following two ways:
\begin{align}
h_1(k) &= v(k) -  E(k) - i(k) \hat R_0; \quad
h_2(k) &= i(k) \hat R_h   \label{eq:h2(k)} 
\end{align}
where 
$v(k)$ is the measured terminal voltage of the battery, 
$i(k)$ is the current through the battery, and 
$E(k)$ denotes the estimate of OCV that is obtained based on the Combined+3 model in \eqref{eq:OCVmodel}.

\section{Data Collection Procedure}
\label{sec:DataCollectionProcedure}

This section details precise data collection procedures for charging the battery and OCV characterization.

\subsection{Fully charging the Battery} 
\label{sec:charging-algorithm}

The battery is charged using a constant-current constant-voltage (CC-CV) approach. 
The term ``fully charged'' can be taken to mean different amounts of Coulomb inputs to the battery based on the threshold current $i_{\rm sd}$ at which the CV charging is terminated.
This is discussed in detail in this section.

\begin{algorithm}
\caption{{CC-CV-Charge}$(i_{\rm cc},i_{\rm sd})$ \label{alg:CCCV}}
\begin{algorithmic}[1]
\State{Measure terminal voltage $v$}
\If{$v<v_{\rm CV1}$}
\State{CC-charge using $i_{\rm cc} $}
\State{Measure terminal voltage $v$}
\If{$v \geq {\rm OCV_{max}} $}
\State{Goto \ref{alg-st:CV}}
\EndIf
\Else
\State{CV-charge at $v = {\rm OCV_{max}}$ \label{alg-st:CV}}
\State{Measure current $i$}
\If{$i<i_{\rm sd}$}
\State{Goto \ref{alg-st:charge-end}}
\EndIf
\EndIf
\State{Terminate Charging \label{alg-st:charge-end}}
\end{algorithmic}
\end{algorithm}

First, the Algorithm \ref{alg:CCCV} describes the CC-CV charging process. 
The CC charging current $i_{\rm cc} $ can take any value less than $i_{\rm max} $ which can be computed as
\begin{align}
i_{\rm max} = \frac{{\rm OCV_{max}}-{\rm OCV_{min}}}{\hat R_0}
\end{align}
where $\hat R_0$ is the estimate of $R_0$ that can be obtained based on approach described in Section \ref{sec:R0-est}. In the absence of an estimate of $R_0$, the rated values provided in the battery data-sheet can be used. 
Typical chargers select the `C-Rate' for the initial charging current $i_{\rm cc} $. 

Initially, the CC-CV charging algorithm checks the voltage $v$ across the battery terminals to decide if the specified current $i_{\rm cc}$ can be safely applied to the battery without triggering the safety protection circuit. 
If the terminal voltage $v$ of a rested battery is greater than 
\begin{align}
v_{\rm CV1} &= {\rm OCV_{max}} -i_{\rm cc} \hat R_0 
\end{align}
it implies that the battery is already sufficiently charged; the charging mode is set to CV instead. 

\begin{figure}[h!]
\begin{center}
\subfloat[][Current vs. Time.]
{\includegraphics[width=0.9\columnwidth]{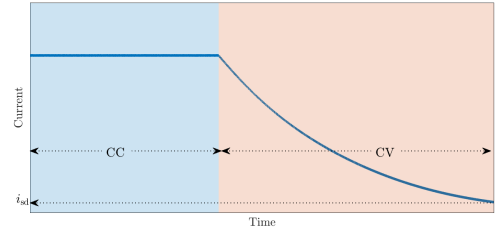}} \\
\subfloat[][Terminal voltage vs. Time.]
{\includegraphics[width=0.9\columnwidth]{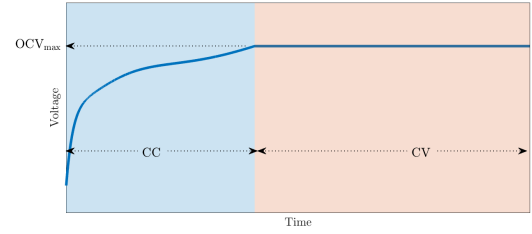}} 
\caption{Current and voltage during the CC-CV data collection.}
\label{fig:VIcccvdata}
\end{center}
\end{figure}

\begin{figure}[h!]
\begin{center}
\includegraphics[width=0.7\columnwidth]{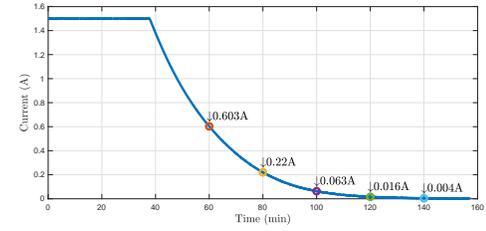}
\caption{{\bf Current during the CC-CV charging for different shutdown thresholds.} For a sample battery, as $i_{\rm sd}$ reduces from $0.603$A to $0.004$A, the charging time increases from $60$ minutes to $140$ minutes.}
\label{fig:CCIsd}
\end{center}
\end{figure}

The charge termination current $i_{\rm sd}$ is an important parameter that determines the true SOC of the battery at the end of the CC-CV charging according to Algorithm \ref{alg:CCCV}. 
The higher the $i_{\rm sd}$ the lower the charging time and final SOC, and vice versa (see Figure \ref{fig:CCIsd}).

A battery can be charged before an OCV test at a $C/N$ rate using one of the following two ways:
\begin{enumerate}
\item
Using the CC-CV strategy presented in Algorithm \ref{alg:CCCV} with the following parameters ($C, C/N$)
\item
Using the CC strategy with a charging current of $C/N$ until the terminal voltage reaches ${\rm OCV_{max}}$
\end{enumerate}
Since the first approach is much faster than the second approach, the first approach is adopted in this paper.
 
 \subsection{OCV Characterization}
\label{sec:ocvchar}

The Algorithm \ref{alg:SlowOCV} summarizes the data collection procedure to obtain open-circuit-voltage parameters. 
Here, the algorithm is named as the `Low-rate OCV test' to contrast it from a recently reported approach in \cite{nguyen2022fast} that proposed a faster approach to OCV characterization. 
In Algorithm \ref{alg:SlowOCV}, the input $N$ refers to $C/N$ rate --- the constant charging current used to perform the OCV characterization data collection. 
It is important to notice that, for the OCV test at the $C/N$ rate, the battery is charged to a charge termination current of $C/N$. 
Once the data is collected, the approaches summarized in Section \ref{sec:OCV-SOC-char} can be used to estimate the OCV parameters.

\begin{algorithm}[h!]
\caption{{Low-rate OCV-Test }$(N, T)$ \label{alg:SlowOCV}}
\begin{algorithmic}[1]
\State{Set Temperature: $T$}
\State{CC-CV-Charge($1C, C/N$)}
\State{1-hour Rest }
\State{CC-discharge $(C/N)$}
\Statex\qquad{Sample data at 1/60 Hz}
\State{CC-charge $(C/N)$}
\Statex\qquad{Sample data at 1/60 Hz}
\State{1-hour Rest}
\State{BattResistanceTest $(1, room, 1 )$}
\end{algorithmic}
\end{algorithm}

\begin{figure}[h!]
\begin{center}
\subfloat[][Current vs. Time.]
{\includegraphics[width=0.9\columnwidth]{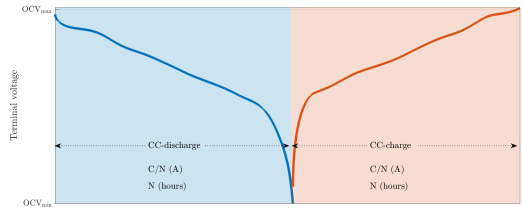}} \\
\subfloat[][Voltage vs. Time.]
{\includegraphics[width=0.9\columnwidth]{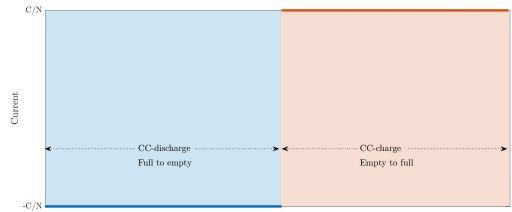}}
\caption{Current and voltage during OCV characterization.}
\label{fig:VIOCVtest}
\end{center}
\end{figure}

\section{Conclusions}
\label{sec:conclusions}

This paper presented systematic approaches to obtaining data for battery characterization required in battery management systems. Particularly, data collection approaches for OCV-SOC characterization and resistance estimation are described in detail. For OCV-SOC characterization, the proposed data collection scheme is designed to reduce modeling error and improve consistency. The low-rate OCV-SOC characterization approach relies on the symmetry of the data during charging and discharging to eliminate the effect of hysteresis in the resulting OCV-SOC model. To achieve such symmetry, it is suggested to charge the battery in a specific way before starting the low-rate data collection for OCV-SOC characterization. In this regard, relevant details about the selection of threshold current that influences the charge input to the battery were explained. 
Additionally, optimal and efficient resistance estimation profiles are described for studying the hysteresis effects on the battery. The data collection was explained for a general C-Rate and can be applied to any battery type. 
In the third part of this series \cite{slowOCVp3}, the data collection procedures summarized in this paper are applied to lithium-ion batteries and the resulting uncertainty on OCV-SOC modeling is analyzed in detail. 


\balance 

\bibliographystyle{ieeetr}
\bibliography{References,literature_BFG}

\end{document}